\documentstyle[prl,aps,psfig]{revtex}
\begin{document}
\draft
\twocolumn[\hsize\textwidth\columnwidth\hsize\csname @twocolumnfalse\endcsname
\title{High-temperature superconductivity in doped antiferromagnets}
\author{Gregory C. Psaltakis}
\address{Department of Physics, University of Crete, and Research Center 
         of Crete, Heraklion, GR-71003, Greece}
\date{Received \today}
\maketitle
\begin{abstract}
In the context of an effective model for doped antiferromagnets, whereby the 
charge carriers are treated as hard-core bosons, we demonstrate that the ground
state energy close to half-filling is an even periodic function of the external 
magnetic flux threading the square lattice in an Aharonov-Bohm geometry.
The period is equal to the flux quantum $\Phi_{0}=2\pi\hbar c/q$ entering the 
Peierls phase factor of the hopping matrix elements. Thus flux quantization and
a concomitant finite value of superfluid weight $D_{s}$ occur along with 
metallic antiferromagnetism. We argue that the charge $q$ in the associated 
flux quantum might be set equal to $2e$. The superconducting transition 
temperature $T_{c}$ is related to $D_{s}$ linearly, in accordance to the 
generic Kosterlitz-Thouless type of transition in a two-dimensional system, 
signalling the coherence of the phase fluctuations of the condensate. 
The calculated dependence of $T_{c}$ on hole concentration is qualitatively 
similar to that observed in the high-temperature superconducting cuprates. 
\end{abstract}

\vspace*{5mm}
\pacs{{\em PACS:} 71.27.+a; 74.20.Mn \\
      {\em Keywords:} High-$T_{c}$ superconductors; Doped antiferromagnets; 
      Flux quantization}
\vskip2pc]

\narrowtext
\section{Introduction}
\label{sec:intro}

The continuous evolution of the charge and spin dynamics upon doping with 
mobile holes the insulating antiferromagnetic copper oxide layers, supports the
view that the high-temperature superconductivity in these materials 
\cite{Bednorz86} is a fundamental property of the two-dimensional doped 
antiferromagnets \cite{Kastner98}. Of particular importance in this respect is
the observed linear increase of the superfluid weight $D_{s}$ with small hole 
concentration $(1-n_{e})$, away from the Mott metal-insulator transition point
at half-filling \cite{Uemura89-93}. On the theoretical side of this problem, 
there has been some evidence that mobile holes in doped antiferromagnets behave
much like hard-core bosons. This transmutation of statistics, from bare 
fermionic holes to bosonic vacancy quasiparticles, should be understood as an 
``emergent phenomenon'' due to the reduced dimensionality and the presence of a
strongly correlated spin background. In the context of the simple fermionic 
$t$-$J$ model, proposed by Anderson \cite{Anderson87-88} to describe such 
systems, the aforementioned evidence comes from exact-diagonalization studies 
of the ground-state energy and the static hole-hole correlation function on 
small clusters \cite{Long92-93,Chen94,Eder97}. Indeed, the possibility of a 
hard-core boson behavior of the charged vacancies in doped antiferromagnets, 
opening the way to Bose-Einstein condensation and the appearance of 
superconductivity, was suggested by many authors \cite{Kivelson87,Thouless87}
in the early days of high-$T_{c}$ superconductivity research. Thouless 
\cite{Thouless87}, in particular, argued that due to topological constraints, 
a vacancy in a two-dimensional torus lattice threaded by an external magnetic 
flux must be transported twice around the ring in order to recover its original 
configuration. Hence flux quantization with an effective charge $q=2e$ may 
result from this period-doubling of the charge $e$ bosons. 

In all the aforementioned works, the lack of an effective model for doped 
antiferromagnets expressed in terms of hard-core bosons has prevented the 
systematic study of their flux quantization properties in conjunction with the 
optical and magnetic ones. Such a model, however, has been postulated from the
outset by Psaltakis and Papanicolaou \cite{Psaltakis93} and consists of a 
$t$-$t^{\prime}$-$J$ Hamiltonian and a suitable $1/N$ expansion that provide 
a reasonably simple many-body calculational framework for the study of the 
relevant issues. When leading quantum-fluctuation effects are taken into 
account in the context of this model, the generic experimental features of the 
optical conductivity, the Drude weight and the total optical weight in the 
cuprates are qualitatively reproduced. In particular, our theory 
\cite{Psaltakis95,Psaltakis96} accounts aptly for the experimentally observed 
$0.5\,\mbox{eV}$ peak of the midinfrared band 
\cite{Cooper90,Orenstein90,Uchida91} and the mass enhancement factor 
approximately equal to 2 \cite{Orenstein90}. Furthermore, it predicts a finite 
limiting value for the optical conductivity $\sigma(\omega\rightarrow 0)$, 
at finite hole doping, consistent with the residual far-infrared conductivity 
observed in the YBa$_{2}$Cu$_{3}$O$_{6+x}$ family of cuprates \cite{Basov95}. 
Our results are also found to be consistent with relevant exact-diagonalization
data \cite{Dagotto94}.

In view of the quoted evidence from optical experiments in favor of our 
effective model, we have recently undertaken a systematic study 
\cite{Psaltakis98} of its flux quantization properties in order to provide a 
more complete assessment of the main electromagnetic responses. Our study 
includes results for the superfluid weight $D_{s}$ and the associated 
superconducting transition temperature $T_{c}$. In particular, our explicit 
numerical estimates for the doping dependence of $T_{c}$, including leading 
quantum-fluctuation effects, are found to reproduce qualitatively the observed 
trends in the cuprates \cite{Uemura89-93,Presland91}. In the following
we review the main points of this approach. 

\section{Effective model}
\label{sec:model}

Our effective model is described by a $t$-$t^{\prime}$-$J$ Hamiltonian 
expressed in terms of Hubbard operators $\chi^{ab}=|a\rangle\langle b|$ as
\begin{equation}
H=-\sum_{i,j}t_{ij}\chi^{0\mu}_{i}\chi^{\mu 0}_{j}
+{\textstyle\frac{1}{2}}J\sum_{\langle i,j\rangle}
(\chi^{\mu\nu}_{i}\chi^{\nu\mu}_{j}-\chi^{\mu\mu}_{i}\chi^{\nu\nu}_{j}) \;,
\label{eq:Hamiltonian}
\end{equation}
where the index 0 corresponds to a hole, the Greek indices $\mu,\nu,\ldots$ 
assume two distinct values, for a spin-up and a spin-down electron, and the 
summation convention is invoked. Here $J$ is the antiferromagnetic 
spin-exchange interaction between nearest-neighbor sites ${\langle i,j\rangle}$
on a square lattice endorsed with periodic boundary conditions and a total
number of sites $\Lambda=\Lambda_{x}\times\Lambda_{y}$, where 
$\Lambda_{x}=\Lambda_{y}$. For the hopping matrix elements $t_{ij}$ we assume 
non-zero values, $t$ and $-t^{\prime}$, only between nearest- and 
next-nearest-neighbor sites, respectively, as dictated by quantum-chemistry 
calculations \cite{Hybertsen90,Sawatzky90} for Cu-O clusters and fits of the 
shape of the Fermi surface observed by angle-resolved photoemission 
spectroscopy \cite{Yu91}. We also generalize the local constraint associated 
with (\ref{eq:Hamiltonian}) to $\chi^{00}_{i}+\chi^{\mu\mu}_{i}=N$, where $N$ 
is an arbitrary integer, and consider the commutation properties of the 
$\chi^{ab}$'s to be those of the generators of the U(3) algebra. 
A representation of the latter algebra in terms of Bose operators can then be 
employed, leading to a generalized Holstein-Primakoff realization that resolves
explicitly the local constraint which gives rise to the hard-core character of 
these bosons. One can then develop a perturbation theory based on the $1/N$ 
expansion, restoring the relevant physical value $N=1$ at the end of the 
calculation. 

In the presence of an external magnetic flux $\Phi$, threading the 
two-dimensional lattice in an Aharonov-Bohm torus geometry, the hopping matrix 
elements $t_{ij}$ are modified by the well-known Peierls phase factor and 
should be substituted in (\ref{eq:Hamiltonian}) according to
\begin{equation}
t_{ij}\leadsto t_{ij}e^{iA_{ij}} \;, \;\;\;
\mbox{with} \;\; A_{ij}=\frac{2\pi\Phi}{\Lambda_{x}\Phi_{0}}
({\bf R}_{i}-{\bf R}_{j})\cdot{\bf e}_{x} \;.
\label{eq:Peierls}
\end{equation}
Here ${\bf R}_{i}$ is the position vector for site $i$, ${\bf e}_{x}$ is the 
unit vector along the $x$-axis encircling the flux lines and 
$\Phi_{0}=2\pi\hbar c/q$ is the so-called flux quantum \cite{comment1}. 
Conventionally, the charge $q$ of the carriers entering $\Phi_{0}$ is, 
of course, equal to the electronic charge $e$. However, the arguments of 
Thouless \cite{Thouless87} quoted in the Introduction imply that a vacancy 
actually ``feels'' twice as much external flux. In the context of the present
{\em effective} model this may be accounted for by an extra factor of two in 
the expression (\ref{eq:Peierls}) for the $A_{ij}$ which can be readily 
absorbed in a redefinition of $q$ as $q=2e$. Evidently, this reasoning does not
constitute a rigorous justification for the assignment $q=2e$ in the flux 
quantum $\Phi_{0}$. The latter justification can be provided only by an 
{\it ab initio} derivation of an effective Hamiltonian for the hard-core boson
vacancies, starting from a realistic electronic model for the cuprates. 
At present such a program is out of reach. Hence this work will be content with
the study of the flux quantization properties of the effective model described 
by (\ref{eq:Hamiltonian})--(\ref{eq:Peierls}), given the flux quantum constant 
$\Phi_{0}$.

In the large-$N$ limit the Bose operators become classical commuting fields 
which can be parametrized by the local electronic density $n_{i}$, the angles
$\theta_{i}$ and $\phi_{i}$ determining the local spin direction, and the local
phase $\psi_{i}$ of the condensate. The Hamiltonian 
(\ref{eq:Hamiltonian})--(\ref{eq:Peierls}) takes the form 
$H(\Phi)=N^{2}\Lambda E_{0}(\Phi)$, where $E_{0}(\Phi)$ is the classical energy
per lattice site for the value of physical interest $N=1$. More explicitly,
for uniform density states, $n_{i}=n_{e}$, where $n_{e}$ is the average 
electronic density, we have that 
\begin{equation}
\Lambda E_{0}(\Phi)={\cal E}_{1}+{\cal E}_{2} \;,
\label{eq:total-energy}
\end{equation}
where
\begin{eqnarray}
{\cal E}_{1} &=& -n_{e}(1-n_{e})\sum_{i,j}t_{ij} 
\nonumber \\
&& \left[\cos\frac{\theta_{i}}{2}\cos\frac{\theta_{j}}{2} 
\cos\left(A_{ij}+\frac{\psi_{i}-\psi_{j}-\phi_{i}+\phi_{j}}{2}\right)\right.
\nonumber \\
&& \left.+\sin\frac{\theta_{i}}{2}\sin\frac{\theta_{j}}{2} 
\cos\left(A_{ij}+\frac{\psi_{i}-\psi_{j}+\phi_{i}-\phi_{j}}{2}\right)\right] \;,
\nonumber \\
&& \label{eq:kinetic-exchange} \\
{\cal E}_{2} &=& \frac{n_{e}^{2}}{4}J\sum_{\langle i,j\rangle}
[\cos\theta_{i}\cos\theta_{j}
+\sin\theta_{i}\sin\theta_{j}\cos(\phi_{i}-\phi_{j})-1] \;.
\nonumber
\end{eqnarray}
As shown in Ref.~\onlinecite{Psaltakis93}, close to half-filling 
($n_{e}\lesssim 1$) and for a sufficiently large $t^{\prime}$, the ground state
of (\ref{eq:total-energy})--(\ref{eq:kinetic-exchange}), in the absence of an 
external magnetic flux ($\Phi=0$), is described by a planar spin configuration 
($\theta_{i}=\pi/2$) in which the local twist angles {\em and} phases are 
modulated according to 
\begin{equation}
\phi_{i}={\bf Q}\cdot{\bf R}_{i} \;, \;\;\;
\psi_{i}={\bf Q}^{\prime}\cdot{\bf R}_{i} \;,
\label{eq:ansatz}
\end{equation}
where ${\bf Q}=(\pi,\pi)$ is the usual spin-modulating antiferromagnetic 
wavevector and ${\bf Q}^{\prime}=(\pi,-\pi)$ is an unusual phase-modulating 
wavevector. The question that is now posed is how this phase-modulated 
antiferromagnetic (AF) ground state will respond to the presence of an external
magnetic flux $\Phi$.

\section{Flux quantization and superfluid weight}

\begin{figure}[h]
\centerline{\psfig{figure=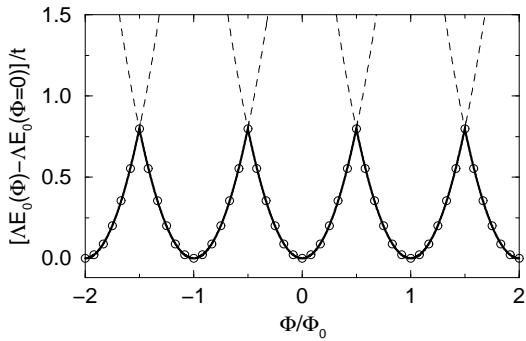,width=7.0cm}}
\caption{\label{fig:AF-flux} Ground-state energy vs external magnetic flux, for
$\varepsilon=0.45$, $t/J=1.0$ and $1-n_{e}=0.10$. The zero-flux energy is 
subtracted to normalize the values. Solid line: the analytic result in the 
infinite lattice limit ($\Lambda\rightarrow\infty$), according to 
Eqs.~(\protect\ref{eq:energy-diff})--(\protect\ref{eq:envelope}). Dashed lines:
remnants of the crossing energy-level parabolas discussed in the text. Open 
circles: numerical minimization results for the ground-state energy on a finite
lattice ($\Lambda=20\times 20$), as determined by 
Eqs.~(\protect\ref{eq:total-energy})--(\protect\ref{eq:kinetic-exchange}). 
Evidently, the finite lattice numerical data (open circles) confirm the infinite
lattice limit analytic result (solid line).}
\end{figure}

Following an argument by Yang \cite{Yang62} we note that, in the presence of 
$\Phi$, the reciprocal lattice is displaced from the origin by 
$2\pi\Phi/(\Lambda_{x}\Phi_{0})$ along the $x$-axis. The quantization of flux 
therefore depends on whether the ground-state energy of the system changes 
under this momentum boost. Given that the spin-exchange part of the Hamiltonian
(\ref{eq:Hamiltonian})--(\ref{eq:Peierls}) does not couple directly to the
magnetic flux it is plausible that, at least in the large-$N$ limit, the 
condensate will respond in such a way as to leave its spin-modulating 
wavevector ${\bf Q}$ intact and simply adjust its phase-modulating wavevector 
${\bf Q}^{\prime}$ to a new value. In other words, we anticipate that in this 
classical (large-$N$) limit, the rigidity of the ground state against the 
intrusion of the external magnetic flux comes solely from the phase 
fluctuations of the condensate. These heuristic arguments lead us to consider 
the {\it ansatz} (\ref{eq:ansatz}) with the following modulating wavevectors:
${\bf Q}=(\pi,\pi)$ and ${\bf Q}^{\prime}=(\pi,-\pi)-(4\pi m/\Lambda_{x},0)$, 
where $m$ is an arbitrary integer. Inserting these wavevestors into 
(\ref{eq:ansatz}) and taking carefully the infinite lattice limit 
($\Lambda\rightarrow\infty$) of the classical Hamiltonian 
(\ref{eq:total-energy})--(\ref{eq:kinetic-exchange}) we have that
\begin{equation}
\Lambda E_{0}(\Phi)-\Lambda E_{0}(\Phi=0)=8t^{\prime}\pi^{2}n_{e}(1-n_{e})
\left(\frac{\Phi}{\Phi_{0}}-m\right)^{2} \;.
\label{eq:energy-diff}
\end{equation}
Thus for each integer $m$ we get an individual many-body energy level that 
depends quadratically on $\Phi$. The ground-state energy is given by the lower 
envelope of these crossing energy-level parabolas and is characterized 
analytically by the condition
\begin{equation}
\left|\frac{\Phi}{\Phi_{0}}-m\right|\leq\frac{1}{2} \;, 
\;\;\; \mbox{with} \;\; m=0,\pm1,\pm2,\ldots \;.
\label{eq:envelope}
\end{equation}
In Fig.~\ref{fig:AF-flux} we depict by a solid line the ground-state energy 
calculated according to (\ref{eq:energy-diff})--(\ref{eq:envelope}), for typical
values of the parameters $\varepsilon=t^{\prime}/t$, $t/J$ and the hole 
concentration $(1-n_{e})$. We also depict by dashed lines the remnants 
of the individual crossing energy levels (\ref{eq:energy-diff}). Evidently,
the ground-state energy (solid line) is an even periodic function of the 
external magnetic flux $\Phi$, with a macroscopic energy barrier between 
different flux minima, in accordance with the Byers and Young \cite{Byers61} 
characterization of a superconductor. The period is equal to $\Phi_{0}$ and
therefore the assignment $q=2e$, discussed earlier on, leads to agreement with 
the observed flux quantization in the high-$T_{c}$ superconducting copper oxide
layers \cite{Gough87}. In order to establish firmly the analytic result 
(\ref{eq:energy-diff})--(\ref{eq:envelope}) we have also minimized numerically 
the classical energy (\ref{eq:total-energy})--(\ref{eq:kinetic-exchange}). 
Excellent agreement was obtained  already for lattices with 
$\Lambda=20\times 20$, as evidenced in the specific example of 
Fig.~\ref{fig:AF-flux}, where the open circles correspond to the numerical 
minimization data.

Let us now turn our attention to the superfluid weight (or helicity modulus) 
$D_{s}$ given by the curvature of the infinite lattice limit of the 
ground-state energy $\Lambda E(\Phi)$ at $\Phi=0$ 
\cite{Byers61,Yang62,Scalapino92-94}, 
\begin{equation}
D_{s}=\Lambda\left(\frac{\Phi_{0}}{2\pi}\right)^{2}
\left[\frac{\partial^{2}E(\Phi)}{\partial\Phi^{2}}\right]_{\Phi=0} \;.
\label{eq:D_s:definition}
\end{equation}
$D_{s}$ determines the ratio of the density of the superfluid charge carriers 
to their mass, and is related to the directly measurable in-plane London 
penetration depth $\lambda_{L}$ by $D_{s}=c^{2}/(4\pi e^{2}\lambda_{L}^{2})$.
Quite generally, $E(\Phi)$ has a $1/N$ expansion of the form
$E(\Phi)=N^{2}E_{0}(\Phi)+NE_{1}(\Phi)+\cdots\;$ which leads via 
(\ref{eq:D_s:definition}) to a corresponding expansion for the superfluid 
weight $D_{s}=N^{2}D_{s}^{(0)}+ND_{s}^{(1)}+\cdots\;$. Hence by exploiting the 
large-$N$ limit result (\ref{eq:energy-diff})--(\ref{eq:envelope}) we get 
immediately the expression for the leading term $D_{s}^{(0)}$,
\begin{equation}
D_{s}^{(0)}=4t^{\prime}n_{e}(1-n_{e}) \;.
\label{eq:Ds(0)}
\end{equation}
Our earlier arguments show that $D_{s}^{(0)}$ is a measure of 
the stiffness of the classical phase fluctuations of the condensate. 
Furthermore, (\ref{eq:Ds(0)}) implies $D_{s}^{(0)}=D_{0}$, where $D_{0}$ is the
leading term in the $1/N$ expansion of the Drude weight 
$D=N^{2}D_{0}+ND_{1}+\cdots\;$, studied in Ref.~\onlinecite{Psaltakis96} using 
the Kubo formalism for the current-current correlations. We have also verified, 
by a straightforward but lengthy calculation of $E_{1}(\Phi)$ and the use of 
(\ref{eq:D_s:definition}), that $D_{s}^{(1)}=D_{1}$. Due to the analytic 
structure of the $1/N$ expansion, these results signify the term-by-term 
validity of the identity $D_{s}=D$. Strictly speaking, of course, we have 
checked explicitly that $D_{s}=D$ only up to and including terms 
$D_{s}^{(1)}=D_{1}$, i.e., only up to and including leading quantum-fluctuation 
effects \cite{comment2}. This, however, is sufficient for most practical 
purposes and permits us to exploit our calculations of the Drude 
weight, in the present study of the superfluid weight. For instance, the weight
$D_{s}=D$, including leading quantum-fluctuation effects, is found 
\cite{Psaltakis96} to increase linearly with small hole concentration 
$(1-n_{e})$ away from the half-filled-band limit ($n_{e}=1$). This trend, 
present already in (\ref{eq:Ds(0)}), is a fundamental characteristic of doped 
antiferromagnets. At higher doping values $D_{s}=D$ eventually saturates and 
then starts to decrease. 

Concerning the expected transition temperature to the charged superfluid, i.e.,
superconducting, state under study we note that at a finite temperature $T$, 
the ratio of the thermal de Broglie wavelength of the charge carriers to their 
average distance is proportional to $\sqrt{D_{s}/(k_{B}T)}$, where $D_{s}$ is 
the zero-temperature value determined by (\ref{eq:D_s:definition}). Hence a 
naive application of the criterion for the occurrence of Bose-Einstein 
condensation in an ideal boson gas, whereby the latter ratio should become of 
order unity, suggests a transition temperature $T_{c}$ of the form
\begin{equation}
k_{B}T_{c}=AD_{s} \;,
\label{eq:Tc}
\end{equation}
where $A$ is a dimensionless constant of order unity. Of course, in the strictly
two-dimensional model of continuous symmetry under study, a {\it bona fide} 
finite temperature phase transition can only be of the Kosterlitz-Thouless type
which, nevertheless, leads again to an expression of the form (\ref{eq:Tc}). 
Indeed, the $\psi_{i}$-structure of the classical Hamiltonian 
(\ref{eq:total-energy})--(\ref{eq:kinetic-exchange}) is a generalization of the
two-dimensional $XY$ model where the latter transition is well studied. In this
context, it is important to note that a ``universal'' linear relation of the 
form (\ref{eq:Tc}) has been established experimentally in the cuprates by 
Uemura {\it et al}. \cite{Uemura89-93} in their remarkable study of $T_{c}$ as
a function of the zero-temperature value of $\lambda_{L}^{-2}\propto D_{s}$. In
the large-$N$ limit, the $D_{s}$ appearing in (\ref{eq:Tc}) is just equal to
$D_{s}^{(0)}$ and the corresponding critical temperature $T_{c}^{(0)}$ should 
be interpreted as the ordering temperature for the classical phase fluctuations
of the condensate, in analogy with the analysis of Emery and Kivelson 
\cite{Emery95} of the classical phase fluctuations of the conventional BCS 
order parameter. The higher order terms in the $1/N$ expansion of $D_{s}=D$ 
capture the effects of the quantum fluctuations and renormalize downwards these
weights \cite{Psaltakis96}, thereby reducing the corresponding value of $T_{c}$.

\begin{figure}[h]
\centerline{\psfig{figure=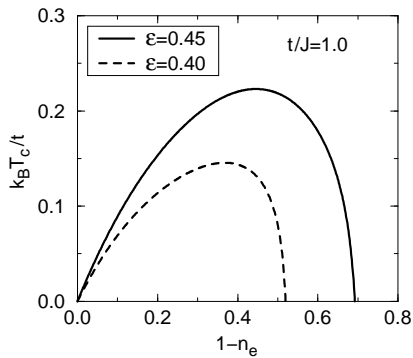,width=5.5cm}}
\caption{\label{fig:hightc} Superconducting transition temperature vs hole 
concentration, for $t/J=1.0$ and $\varepsilon=0.45$ (solid line) or 
$\varepsilon=0.40$ (dashed line), according to Eq.~(\protect\ref{eq:Tc}) 
with the inclusion of leading quantum-fluctuation effects.}
\end{figure}

Following the prescription of Emery and Kivelson \cite{Emery95}, we have 
applied (\ref{eq:Tc}) with $A=0.9$; a numerical value extracted from the 
two-dimensional $XY$ model \cite{Gupta88}. Using the calculated $D_{s}=D$ 
of Ref.~\onlinecite{Psaltakis96}, with the inclusion of the leading 
quantum-fluctuation correction $D_{s}^{(1)}=D_{1}$, we depict in 
Fig.~\ref{fig:hightc} the superconducting transition temperature $T_{c}$ as a 
function of the hole concentration $(1-n_{e})$. Evidently, the dependence of 
$T_{c}$ on $(1-n_{e})$ reflects that of $D_{s}$ and reproduces qualitatively 
the observed trends in the cuprates \cite{Uemura89-93,Presland91}. In 
particular, we have that: $T_{c}\propto (1-n_{e})$, for $n_{e}\rightarrow 1$. 
With an estimated $J/k_{B}\approx 1500\,\mbox{K}$ in the cuprates 
\cite{Shamoto93}, the value of $T_{c}$ at optimum doping $(1-n_{e})=0.44$ 
($0.36$), seen in the solid (dashed) line of Fig.~\ref{fig:hightc}, 
is $T_{c}\approx 335\,\mbox{K}$ ($218\,\mbox{K}$). This predicted value of 
$T_{c}$, signalling the coherence of the phase fluctuations of the condensate, 
should be regarded as an upper bound to an actual transition temperature 
because of the neglect of impurity disorder, higher-order quantum fluctuations,
etc. From Fig.~\ref{fig:hightc} we also note that with further hole doping 
$T_{c}$ starts to decrease while beyond a critical doping value it vanishes, 
as the phase-modulated AF configuration, around which the present $1/N$ 
expansion is carried out, becomes unstable.

\section{Conclusions}

We have demonstrated that flux quantization and a concomitant finite value of 
superfluid weight $D_{s}$ occur in the metallic phase-modulated AF ground state
of the $t$-$t^{\prime}$-$J$ model (\ref{eq:Hamiltonian}). By appealing to the 
universality class of the two-dimensional $XY$ model, the corresponding 
superconducting transition temperature $T_{c}$ is related to $D_{s}$ linearly, 
via (\ref{eq:Tc}). The inclusion of leading quantum-fluctuation effects in 
$D_{s}$ provides then a reasonable estimate for the order of magnitude and the 
doping dependence of $T_{c}$ in the cuprates. These results support our 
effective description of the charge carriers in terms of hard-core bosons.

\section*{Acknowledgments}

The author would like to thank X. Zotos, E. Manousakis, and G. Varelogiannis 
for stimulating discussions. This work was supported by grant No. 
$\Pi$ENE$\Delta$95-145 from the Greek Secretariat for Research and Technology.

\begin{thebibliography}{99}
\bibitem{comment2} Note that for a conventional BCS superconductor with a finite
energy gap, a case radically different from the one under study, the identity 
$D_{s}=D$ is proved in Ref.~\onlinecite{Scalapino92-94}.
\end{thebibliography}
\end{document}